\documentstyle[11pt]{article}
\newlength{\breite}
\newcommand{\ab}[1]{\settowidth{\breite}{$#1$} \mbox{\hspace{\breite}}}

\newcommand{\half}{{\textstyle\frac{1}{2}}}
\newcommand{\fourth}{{\textstyle\frac{1}{4}}}

\newcommand{\nn}{\nonumber}

\newcommand{\dolk}{\mbox{$ / \!\!\! k$}}
\newcommand{\dolp}{\mbox{$ / \!\!\! p$}}

\newcommand{\tr}{\hbox{tr}}
\newcommand{\Tr}{\hbox{Tr}}

\newcommand{\bays}{\left\{\begin{array}{c}}
\newcommand{\eays}{\end{array}\right\}}
\newcommand{\beq}{\begin{eqnarray}}
\newcommand{\eeq}{\end{eqnarray}}

\begin{document}
\def\slash#1{#1 \hskip -0.5em / }

\begin{titlepage}
\thispagestyle{empty}
\begin{flushright}
UNIT\"U-THEP-1/1995\\
hep-ph/9501333\\
January 1995
\end{flushright}
\vspace{0.3cm}
\begin{center}
\Large \bf $\rho\omega$--Mixing and the Pion Electromagnetic Form Factor
        in the Nambu--Jona-Lasinio Model
 \\
\end{center}
\vspace{0.5cm}
\begin{center}
R.\ Friedrich$^{\hbox{\footnotesize{1,2}}}$ and H.\
Reinhardt$^{\hbox{\footnotesize{1}}}$\\
{\sl Institut f\"ur Theoretische Physik, Universit\"at T\"ubingen, \\
  Auf der Morgenstelle 14, D-72076 T\"ubingen, Germany}
\end{center}
\vspace{0.6cm}

\begin{abstract}
\noindent
The $\rho\omega$--mixing generated by the
isospin breaking of the current quark masses $m_{u} \neq m_{d}$
 is studied within the
bosonized NJL model in the gradient expansion. The resulting
effective meson lagrangian naturally incorperates vector meson
dominance. By including pion loops an excellent description of both
the pion electromagnetic form factor and of the $\pi^+\pi^-$ phase shifts
in the vector--isovector channel is obtained. The $\rho\omega$--mixing
can be treated in the static approximation but is
absolutely necessary to reproduce the fine structure of the
electromagnetic form factor, while the pion loops are necessary
to obtain the correct energy dependence of the phase shifts.
\end{abstract}
\bigskip\indent
\setcounter{footnote}{1}
\footnotetext{Supported in part by
 {\it COSY} under contract 41170833}
\setcounter{footnote}{2}
\footnotetext{ e--mail: friedric@ptdec5.tphys.physik.uni-tuebingen.de}
\vfill
\thispagestyle{empty}
\end{titlepage}

\section{Introduction}
\setcounter{page}{1}
\bigskip\indent

Since the success of current algebra it has been known that the
low energy meson dynamics is dominated by chiral symmetry and
its spontaneous and explicit breaking. This is the conceptual
basis for chiral perturbation theory \cite{GL84} where the a priori unknown
effective meson theory underlying QCD is systematically expanded
in powers of derivatives of the meson fields and the arising
expansion coefficients are considered as phenomenological
parameters to be determined from the data.

The mechanism of
spontaneous breaking of chiral symmetry with the appearance of the
pseudoscalar mesons as Goldstone bosons has been well understood
in a microscopic fashion within the bosonized Nambu--Jona-Lasinio
model \cite{ER86}. The gradient expansion of the bosonized NJL model
essentially reproduces the chiral perturbation expansion and
expresses the various expansions coefficients of chiral
perturbation theory in terms of a few microscopic parameters of the NJL
model. Furthermore the bosonized NJL model also shows how the
explicit symmetry breaking by the small current quark masses
gives rise to finite masses of the pseudoscalar mesons.
The current quark masses do explicitly break not only chiral
symmetry but in addition also charge or isospin symmetry. This
breaking of charge symmetry induces various meson mixings which
have recently received increasing interest. For a comprehensive
review see for example \cite{MNS90}.

Given the success of the bosonized NJL model in the isospin
symmetric case one can expect that this model also accounts for
the isospin symmetry breaking effects like the $\rho\omega$--mixing,
since the explicit flavor symmetry breaking through
the current quark masses is included in this model. Some attempts
in this direction have been undertaken in \cite{Vo86}
where the NJL model with a sharp ultraviolet momentum cutoff
was studied in lowest order gradient expansion. However a sharp
cutoff spoils gauge invariance, which becomes important when
electromagnetic processes are considered, and as a consequence
the Ward identities are violated and electric charge is not
conserved.

So far the bosonized NJL model has been mainly used in the
gradient expansion on tree level, which has proved already
sufficient to reproduce the gross feature of the low energy
meson data. In this paper we want to push the model even further
and study processes where on the one hand one has to go beyond
the tree approximation and include pion loops and on the other
hand has to include the subtle effects of meson mixings induced
by the small charge symmetry breaking of the current quark mass. For
this purpose we shall
study the electromagnetic form factor of the pion and the $\pi^+\pi^-$
phase shifts in the vector--isovector channel. We will show
that both the form factor and the phase shifts can be quite
satisfactory reproduced within the bosonized NJL model in
leading order gradient expansion provided one includes the $\rho
\omega$--mixing and pion loops. For a quantitative analysis it is
important to describe both processes simultaneously. According
to vector meson dominance\footnote{The bosonized NJL model naturally
accounts for vector meson dominance \cite{ER86}.}
 they are both determined by the
$\rho$--propagator, which is dressed by pion loops and modified by the
$\rho\omega$--mixing. While the pion loops are necessary to
reproduce the correct energy dependence of the phase shifts the
$\rho\omega$--mixing will show up in the fine structure of the
electromagnetic form factor.

The organization of the paper is as follows:
In the next section we define the NJL model and present the
effective meson lagrangian resulting in leading order gradient
expansion after bosonization. In particular we derive the $\rho
\omega$--mixing induced by the charge symmetry breaking of the
current quark masses. In section 3 we calculate the effective
$\rho$
meson propagator which contains besides the $\rho
\omega$--mixing also pion loops. Section 4 is devoted to the numerical
results.

\section{The NJL Model and the effective meson lagrangian}

The NJL model \cite{NJL61} was originally introduced as an effective
theory for the nucleon field. Subsequently, it had a revival as a
model for the low--energy quark flavor dynamics. The two--flavor NJL
model is defined by the following lagrange density
\beq
{\cal L}
   &=& \overline{q} (i \slash{\partial} - \widehat{m}_0)q +
   \frac{G_1}{2} \left( (\overline{q} \tau^a q)^2 + (\overline{q}
   i\gamma_5 \tau^a q)^2 \right)
\nonumber\\
&&\ab{\overline{q} (i \slash{\partial} - \widehat{m}_0)q} - \frac{G_2}{2}
	\left(
      (\overline{q} \gamma^\mu \tau^a q)^2
      + (\overline{q} \gamma^\mu \gamma_5 \tau^a q)^2
	\right)\quad .
\label{njl}
\eeq
Here $q=(u,d)^T$ denotes the quark field, $\tau^a$ are the isospin
Pauli matrices, $\widehat{m}_0 = diag\left(m_0^u,m_d^0\right)$ is the
current quark matrix and $G_1$ and $G_2$ are coupling constants of
dimension $({\rm MeV})^{-2}$. For $\widehat{m}_0 = 0$ the model has
${\rm SU(2)}_R \times {\rm SU(2)}_L$ chiral symmetry. Note that chiral
symmetry allows for independent coupling constants in the
scalar--pseudoscalar and vector--axialvector sectors.\footnote{If the
interaction in (\ref{njl}) is deduced from one gluon exchange by Fierz
transformation one obtains $G_2= \half G_1$.}

The generating functional for Green's functions of quark
bilinears is given by
the path integral
\begin{equation}
Z(\eta)= \int {\cal D} q {\cal D} \overline{q}
e^{i \int d^4x (
{\cal L}(q) + {\cal L}_{source}(j))}
\quad .
\label{action}
\end{equation}
Here we have included a source term
\begin{eqnarray}
  {\cal L}_{source}(j) &=&
  j (\overline{q} \Gamma q) + h.c.
\label{source}
\quad ,
\end{eqnarray}
where $\Gamma$ is a suitable combination of Dirac and isospin matrices.

Applying by now  standard bosonization techniques \cite{ER86} one
converts the NJL model (\ref{njl}) into an effective meson theory, which can
be cast into the following form \cite{RD89}

\begin{eqnarray}
\int {\cal L} &=&
   -i\Tr \ln\left(i \slash{\partial}
	- \Sigma
        + \slash{V} + \slash{A}\gamma_5 \right)
\nonumber \\
&&-\frac{1}{4G_1} {\rm tr_F} \int
 \left[ (\Sigma)^2 -
 \widehat{m}_0 (\xi_L^\dagger \Sigma \xi_R + \xi_R^\dagger \Sigma \xi_L)
 \right]
\nonumber\\
&&+\frac{1}{4G_2} {\rm tr_F}\int
 \left[ (V_\mu - {\cal V}_\mu^\pi)^2
       +(A_\mu - {\cal A}_\mu^\pi)^2
 \right] + S_{anom}
\quad
\label{gauss}
\end{eqnarray}
Here $V_\mu=V_\mu^a\tau^a$ and $A_\mu=A_\mu^a\tau^a$
denote the vector and axialvector fields, respectively. Furthermore $\Sigma$
is a scalar field (the so called chiral radius) and $\xi_{L,R}$ are
unitary fields, which are related to the more standard scalar $s= s^a
\tau^a$ and pseudoscalar $p=p^a\tau^a$ fields by
\begin{equation}
 s + i  p = \xi_L^\dagger \Sigma \xi_R
\quad .
\end{equation}
Moreover we have introduced the induced vector and axialvector fields
\begin{eqnarray}
  {\cal V}_\mu^\pi &=&
\frac{i}{2} (\xi_R \partial_\mu \xi_R^\dagger +
             \xi_L\partial_\mu \xi_L^\dagger)  \nonumber \\
  {\cal A}_\mu^\pi &=&
\frac{i}{2} (\xi_R \partial_\mu \xi_R^\dagger -
             \xi_L\partial_\mu \xi_L^\dagger)\nonumber\quad ,
\end{eqnarray}
which originate from a chiral rotation of the original quark fields
(see ref. \cite{ER86}). The corresponding Jacobian \cite{Fu80} yields
the (integrated) chiral anomaly $S_{anom}$, the explicit form of which
we do not need in the following. (In leading order gradient expansion
$S_{anom}$ is given by the Wess--Zumino action \cite{WZ71}.)
Under a chiral transformation ${\rm SU(2)}_R \times {\rm SU(2)}_L$ of
the quarks
\beq
q_L\to L q_L\qquad q_R\to R q_R\nn
\eeq
the meson fields transform according to
\begin{eqnarray}
&&  \xi_R(x) \to h(x) \xi_R (x) R^\dagger \quad , \qquad
  \xi_L(x) \to h(x) \xi_L (x) L^\dagger\quad , \nn\\
&& \Sigma \to h(x) \Sigma h^\dagger(x)\quad , \nn\\
&&V^\mu \to h(x) V^\mu h^\dagger(x) + i h(x)\partial^\mu
h^\dagger(x)\quad ,\nn \\
&&A^\mu \to h(x) A^\mu h^\dagger(x)
\quad , \nonumber
\end{eqnarray}
where $h(x)$ is an element of the so called local hidden symmetry group
\cite{BKY88}. Due to this symmetry the effective action (\ref{gauss})
depends on $\xi_{L,R}$ only via the chiral field
\beq
U(x) = \xi^\dagger_L(x)\xi_R(x)\nn\quad .
\eeq
To remove this extra gauge symmetry we adopt the unitary gauge
\begin{equation}
\xi_R=\xi_L^\dagger=\xi\quad .
\label{unigau}
\end{equation}
The quark loops is a diverging object and needs regularization. We
will use the proper time regularization throughout the paper.

In the vacuum defined by the stationary points of the effective
action, only the scalar field $\Sigma$ develops a nonzero
expectation value

\begin{equation}
\Sigma = \hat{m} = diag(m_u, m_d)\quad ,
\end{equation}
which represents the constituent quark mass $m_i$, $i=u,d$ and signals the
spontaneous breakdown of chiral symmetry.
Stationarity of the action with respect to variation in $\Sigma$
yields the following gap equation

\begin{equation}
m_i = m_i^0 +  G_1  \frac{N_c}{2\pi^2}  \left(m^i\right)^3
\Gamma(-1,(m^i)^2/\Lambda^2)\quad ,
\end{equation}
\bigskip

\noindent
whose solutions yield flavor dependent constituent quark masses
$m_i, i = u,d$. Hence the vacuum configuration $m_i, i = u,d$
breaks already charge symmetry.

The physical mesons are given by the small amplitude excitations
of the meson fields around their vacuum value. We shall
concentrate on the low lying non--strange mesons, which are the
$\pi ,\rho$ and $\omega$, and put the scalar field $\Sigma$ on its
vacuum value. The contact terms in the effective meson action
will give rise to the meson masses. The quark loop depends, with
$\Sigma$ fixed at its vacuum value, only on the vector and axialvector fields
$V_\mu$ and $A_\mu$.
Expansion of the quark loop in powers of
$V_\mu$ and $A_\mu$ yields in second order:
\beq
-i\Tr \ln\left(i \slash{\partial}
	- \Sigma
        + \slash{V} + \slash{A}\gamma_5 \right)
&=& -i\Tr \ln\left(i \slash{\partial}
	- \Sigma \right)\nn\\
& & +\half\int V^\mu_{a}(x)\Pi_{\mu\nu}^{V\;ab}(x,y)V^\nu_b(y)\nn\\
& & +\half\int A^\mu_{a}(x)\Pi_{\mu\nu}^{A\;ab}(x,y)A^\nu_b(y)
\eeq
where the meson self energy is given by

\beq
\Pi_{\mu\nu}^{(V,A)\;ab}(x,y) = \frac{-i}{2} \Tr\left(
G_0(y,x)\gamma_\mu \bays 1 \\ \gamma_5\eays\tau^a
G_0(x,y)\gamma_\nu \bays 1 \\ \gamma_5\eays\tau^b
\right)
\label{self1}
\eeq

Following \cite{ER86} we will perform a gradient expansion

\beq
\Pi_{\mu\nu}^{(V,A)\;ab}(p^2) =\Pi_{\mu\nu}^{(V,A)\;ab}(0)
+\left(\frac{d}{dp^2}\Pi_{\mu\nu}^{(V,A)\;ab}(p^2)\right)_{p^2=0} + \dots
\eeq

For the charge symmetric case extraction of the meson
properties in the gradient expansion of the effective meson
theory has been performed in \cite{ER86} . After redefinition
of the meson fields to eliminate the $\pi a_1$--mixing and to
bring the resulting effective meson lagrangian in the standard
form one finds the following results \cite{RD89}

\beq
  M_\pi^2 &=& \frac{m_0 m}{G_1 F_\pi^2} \quad , \qquad
  M_V^2 = \frac{g_V^2}{4G_2} \quad , \qquad
  M_A^2 = M_V^2 + 6m^2  \quad , \nonumber\\
  F_\pi^2 &=&\frac{1}{4G_2}\left(1-\frac{M_V^2}{M_A^2} \right)
  \quad , \qquad
  g_{V\pi\pi}  =  \frac{g_V}{8G_2 F_\pi^2} \quad ,
\eeq
with
\beq
g_V^{-2} =
\frac{N_c}{24\pi^2}\Gamma\left(0,\frac{\Sigma_0^2}{\Lambda^2}\right) \qquad
\Sigma_0=\half (m_u+m_d)\quad .
\eeq
Here we shall go beyond \cite{ER86,RD89} and include in addition
the charge symmetry breaking which gives rise to the $\rho
\omega$--mixing. As well as the dressing of the $\rho$ propagator
by pion loops. The analogous $a_1a_D$--mixing is experimentally
less understood. For this purpose we ignore the $a_1$ channel in
the following (after the $\pi a_1$--mixing has been properly removed).
When the isospin symmetry breaking part $\Delta \Sigma$ defined by

\beq
\Sigma = \Sigma_0\tau_0 + \Delta \Sigma\tau_3\qquad ,
\Sigma_0 = \half(m_u &+&m_d)\qquad ,\Delta \Sigma= \half(m_u-m_d)
\eeq
is included the vector meson self energy $\Pi^V$ contains besides the flavor
diagonal parts considered already in \cite{ER86} also an
isospin mixing part $\Pi_{\mu \nu}^{V\,03}(p^2)$ which gives rise to a
$\rho\omega$--mixing. Note also that the vector meson mass terms
do not lead to any $\rho\omega$--mixing. So the total mixing ${\cal
L}_{\rho\omega}$ comes entirely from the quark loop
\beq
\int {\cal L}_{\rho\omega}= \int \rho_\mu^3\Pi_{\mu\nu}^{V\,03} \omega_\nu
\eeq

 From Lorentz invariance we expect the mixing term to be of the
form (in momentum representation)
\beq
{\cal L}_{\rho\omega} = -m_{\rho\omega}^2(p^2) \rho^3_\mu\omega^\mu +
                 f(p^2) p^\mu \omega_\mu p^\nu \rho^3_\nu \quad .
\label{lmix}
\eeq
The last term contains the derivative couplings of $\rho$ and
$\omega$. This term will not contribute to the electromagnetic
form factor of the pions where the $\rho$ couples to the
conserved electromagnetic current. In fact by vector meson dominance
we have  $j_\mu^{em} \sim \rho_\mu$ and $\partial^\mu j_\mu^{em} = 0$
implies $\partial^\mu  \rho_\mu = 0$.

 From equation (\ref{self1}) the mixing term can be straight forwardly
evaluated. Working out the isospin trace shows that $\Pi^{V\,03}$
is given by the difference between the $u$ and
$d$ vector quark loops
\beq
\Pi_{\mu\nu}^{V\,03}(x,y) = \frac{-i}{2}\Tr\;\left( G_u(x,y)\gamma_\mu
G_u(y,x)\gamma_\nu - (u\to d)\right)
\eeq
Using here for the quark loop again the proper time regularization
one finds
\beq
-m_{\rho\omega}^2\rho^3_\mu\omega^\mu + f p^\mu\rho^3_\mu p^\nu\omega_\nu &=& i
N_c
\int \frac{d^4k}{(2\pi)^4} {\rm tr} \frac{\dolk -\dolp + m_u}{(k-p)^2
-m_u^2}\slash{\rho}^3 \frac{\dolk + m_u}{k^2-m_u^2}\slash{\omega} \quad-\quad
\nonumber\\& &\left(m_u\rightarrow m_d\right)\nonumber\\
&=&\frac{N_c}{2\pi^2}
\left(p^2\rho^3_\mu\omega^\mu- p^\mu\rho^3_\mu p^\nu\omega_\nu\right)
\int_0^1 dx x (1-x)\times\nn\\ & &\!\!\!\!\!\! \left\{
\Gamma\left(0,\frac{(m_u^2-x(1-x)p^2)}{\Lambda^2} \right)-
\Gamma\left(0,\frac{(m_d^2-x(1-x)p^2)}{\Lambda^2} \right)
\right\} ,
\label{mix2}
\eeq
from which we identify the $\rho\omega$--mixing
\beq
m^2_{\rho\omega}(p^2) =2 p^2\frac{\int_0^1 dx x (1-x)
\Gamma\left(0,\frac{m_d^2-x(1-x)p^2}{\Lambda^2} \right) -
\Gamma\left(0,\frac{m_u^2-x(1-x)p^2}{\Lambda^2} \right)}{
\int_0^1 dx x (1-x)\Gamma\left(0,\frac{\Sigma_0^2-x(1-x)p^2}{\Lambda^2} \right)
}
\quad .
\label{mrw}
\eeq
Let us emphasize that
\beq
m_{\rho\omega}^2\left(p^2=0\right) = 0\quad ,
\label{zero}
\eeq
which is a consequence of the gauge symmetry
preserving regularization method used. If we had
used another regularization scheme, e.g. a sharp Euclidean cutoff, we
would have got also a term which does not vanish for $p^2 = 0$ \cite{Vo86}.
The gradient expansion of the $\rho\omega$--mixing yields in
leading order
\beq
m^2_{\rho\omega}(p^2) =2 p^2\frac{
\Gamma\left(0,\frac{m_d^2}{\Lambda^2} \right) -
\Gamma\left(0,\frac{m_u^2}{\Lambda^2} \right)}{
\Gamma\left(0,\frac{\Sigma_0^2}{\Lambda^2} \right) }\quad .
\label{mrw2}
\eeq

On the mass shell $p^2 = M_\rho^2 \approx M_\omega^2 \approx
0.6\;{\rm GeV}^2$ this mixing strength is empirically known to be
$m_{\rho \omega}^2 = (4520 \pm 600) {\rm MeV}^2$ . Fixing all parameters of the
NJL model except for $m_d$ in the isospin symmetric
sector (see \cite{ER86}) the empirically value requires $m_d -
m_u \approx (1-2) {\rm MeV}$ for $m_u = 300 {\rm MeV}$.

In the bosonized NJL model all quark observables become
functionals of the meson fields. It is straight forward to derive the
corresponding expressions for the quark currents by including
appropriate sources $j$ according to (\ref{source}) in the original
quark theory (\ref{action}) and taking
at the end derivatives with respect to the sources in the
bosonized theory. For the electromagnetic current one finds
\beq
j_\mu^{elm} = \frac{M_\rho^2}{g_V}\rho_\mu^3 +\frac{1}{3}
\frac{M_\omega^2}{g_V}\omega_\mu +
\frac{\sqrt{2}}{3}\frac{M_\Phi^2}{g_\Phi}\Phi_\mu \quad ,
\label{elm}
\eeq
which is a manifestation of the vector meson dominance hypothesis
according to which the photon couples to hadrons via the vector
mesons. The vanishing of the $\rho\omega$--mixing at $p^2 = 0$
(see equation (\ref{zero})) implies that an onshell photon can couple
to the pion only via the $\rho$--meson since only the $\rho$--meson
couples in the effective meson lagrangian (\ref{gauss}) to the pionic
vector current ${\cal V}^\mu$.

\section{ The effective $\rho $ propagator}

Using the leading order gradient expansion and ignoring the
axial vector mesons (after its mixing with the $\pi$--field is
removed) the effective meson lagrangian obtained in the previous
section by expanding the quark loop up to the second order in the
vector meson and pion fields is given by
\beq
{\cal L} = {\cal L}_0^\rho + {\cal L}_0^\pi + {\cal L}_0^\omega +
           {\cal L}_{\rho\omega}  + {\cal L}_{\rho\pi}
\quad ,
\label{mini}
\eeq
Here
\beq
{\cal L}_0^\rho &=&-\fourth
\left(\partial_\mu\rho^a_\nu -\partial_\nu\rho^a_\mu \right)^2 +\half
m_\rho^2 \rho_\mu^{a\,2}\quad ,\\
{\cal L}_0^\omega &=&-\fourth
\left(\partial_\mu\omega_\nu -\partial_\nu\omega_\mu \right)^2 +\half
m_\omega^2 \omega_\mu^2\quad ,\\
{\cal L}_0^\pi &=& \half \left(\partial_\mu\pi^a\right)^2-\half m_\pi^2
\pi^{a\,2} \quad .
\label{lag}
\eeq
are the free lagrangians of the $\rho$, $ \omega$  and $\pi$ and
\beq
{\cal L}_{\rho\omega} =
m^2_{\rho\omega}\left(p^2 = m_\omega^2\right)\rho^3_\mu\omega^\mu \quad .
\eeq
is the $\rho \omega$--mixing, which has been taken on the mass shell
$p^2 = m_\omega^2$ and
\beq
{\cal L}_{\rho\pi} =  i g_{\rho\pi\pi} \epsilon^{abc}\rho^a_\mu
           \pi^b \partial^\mu\pi^c
\eeq
arises from the expansion of the vector current in (\ref{gauss})
in leading order in the pion field.\footnote{One could as
well keep the full momentum dependence of the $\rho\omega$--mixing.
This would however yield almost identical results due to
the weakness of the $\rho\omega$ mixing, which matters only at
the $\rho$--pole.} We have ignored here the contribution from the chiral
anomaly $S_{anom}$. The Wess--Zumino--action
yields the leading term where $\omega_\mu$ couples to the topological current
$B_\mu$
\beq
\int \omega_\mu B^\mu \quad , B^\mu =
\frac{1}{24\pi^2}\epsilon^{\mu\nu\kappa\lambda} \tr L_\nu L_\kappa
L_\lambda \sim \epsilon^{\mu\nu\kappa\lambda}
\epsilon^{abc}\partial_\nu\pi^a \partial_\kappa \pi^b\partial_\lambda
\pi^c\quad .
\eeq
This term would contribute to the $\omega$--propagator only
through two--loop pion diagrams. In accord with the counting of
chiral perturbation theory where vector meson loops are subleading
to pion loops we will treat the effective meson lagrangian in
tree approximation concerning the vector mesons but include the
one pion loops. Thereby we will concentrate on the
electromagnetic form factor of the pion and the $\pi^+\pi^-$ phase
shifts in the vector--isovector channel, which both are
exclusively determined by the $\rho$--propagator. The electromagnetic form
factor of the pion $F_{em} (q^2)$ is directly related to the
propagator of the $\rho$--meson $D_\rho(k^2)$ by
\beq
\label{ffem}|F_{em}(k^2)|\propto |D_\rho(k^2)|\; ,
\eeq
which shows that the form factor directly probes the momentum
dependence of the $\rho$--propagator. The missing normalization
factor is determined by charge conservation
\beq
 F_{em}(k^2=0)= 1\; .
\eeq
While the electromagnetic form factor obviously measures only
the module of $D_\rho(k^2)$ the phase of the propagator (without
$\rho\omega$--mixing)  is
measured by the pion phase shifts in the vector--isovector channel.
For the scattering amplitude $a_1^1$ one has
\beq  a^1_1 \propto \exp (i \delta ^1_1) \sin (\delta ^1_1) \propto D_\rho\; ,
\eeq
which yields for the phase shift $\delta_1^1$
\beq \tan (\delta ^1_1) =
-\frac{\rm{Im}D_\rho^{-1}}{\rm{Re}D_\rho^{-1}}\; .
\label{ps}
\eeq

%
%begin FigureOB 1
%
\begin{figure}[hbt]
%\begin{center}
%\include{diarho}
%\end{center}
\caption[]{The dressing of the $\rho$--propagator by pion loops.
The curly lines denote the dressed $\rho$--propagator (thick curly
line) and the free $rho$--propagator (thin curly line), respectively.
}
\label{fig1}
\end{figure}
%
%end Figure 1
%

In following we therefore calculate explicitly the $\rho$--propagator.
We include the one pion
loop to the $\rho$--propagator as shown in figure (\ref{fig1}). It
gives rise to a momentum dependent self energy of the $\rho$--meson
and provides a finite width:
\beq
\Sigma^{\mu\nu}_\rho(p) &=& \frac{ig_{\rho\pi\pi}^2}{2}
\int\frac{d^4k}{(2\pi)^4}
\frac{\left(2p-k\right)^\mu}{(k-p)^2-m_\pi^2}
\frac{\left(2p-k\right)^\nu}{k^2-m_\pi^2}\nonumber\\
&=& \frac{g^2}{16\pi^2}\int_0^1 dx
\left\{2\delta^{\mu\nu} \left(m_\pi^2-x(1-x)p^2\right)
\Gamma\left(-1,\frac{m_\pi^2-x(1-x)p^2}{\Lambda_\pi^2}\right) -
\nonumber\right.\\& &\ab{\frac{g^2}{16\pi^2}\delta{ab}\int_0^1 dx}\left.
p^\mu p^\nu (1-2x)^2
\Gamma\left(0,\frac{m_\pi^2-x(1-x)p^2}{\Lambda_\pi^2}\right)
\right\}
\quad .
\label{self2}
\eeq
Here $\Lambda_\pi$ is a new cutoff which chops off the high
momenta of the pion loop. This cutoff which is related to the
size of the meson is independent of the quark loop cutoff
$\Lambda$, which is the scale of spontaneous breaking of chiral
symmetry. Again $\Lambda_\pi$ indicates the range of validity of
the effective meson theory. Since the effective meson theory
arose from the bosonized NJL model in the gradient expansion we
expect that $\Lambda_\pi < \Lambda$. This will come out later
from the actual calculations. Let us also note that the
self energy $\Sigma_\rho$ is not transversal. This should come with
no surprise since the $\rho$--meson does not couple to a
conserved current. Nevertheless the longitudinal part will
contribute neither to the electromagnetic pion form factor
(since the electromagnetic current is conserved) nor to the $\pi^+
\pi^-$  phase shifts (due to the kinematical structure
of the $\rho\pi\pi$ vertex).

%
%begin FigureOB 2
%
\begin{figure}[hbt]
%\begin{center}
%\include{ffmom}
%\end{center}
\caption[]{The electromagnetic form factor of the pion. The photon
couples according to vector meson dominance to the $\rho$--meson and
due to the $\rho\omega$--mixing also to the $\omega$--meson.
The $\rho$--propagator is dressed by pion loops whereas the
$\omega$--propagator is given by a simple Breit--Wigner form. We neglected
the error bars of the data points (represented by small circles) in
order to show the quality of the
fit. The experimental data are taken from \cite{Am84}.
}
\label{fig2}
\end{figure}
%
%end Figure 2
%

%
%begin Figure 3
%
\begin{figure}[hbt]
%\begin{center}
%\include{mix}
%\end{center}
\caption[]{The momentum dependence of the lowest order gradient
expansion of the $\rho\omega$--mixing $m_{\rho\omega}^2$ for three
different constituent quark masses $m_d$. All other parameters are
chosen to reproduce the empirical values for the low energy
observables $F_\pi, m_\pi, m_\rho$ and $g_{\rho\pi\pi}$.
Then $m_d$ remains as free parameter while
$m_u = 300\; {\rm MeV}$ is fixed.
The bold lines shows $m_{\rho\omega}^2$ for $m_d = 301\; {\rm MeV}$, the
thin line for $m_d = 302\; {\rm MeV}$ and the dashed line is calculated
with $m_d = 303\; {\rm MeV}$. For comparison, $m_u$ has $300\; {\rm MeV}$.
The empirical value for $m_{\rho\omega}^2$ is only known on the vector
meson mass shell $p^2\approx\left(m_\rho^2+m_\omega^2\right)/2\approx
0.6\;{\rm GeV}^2$ being $m_{\rho\omega}^2\approx \left(-4.5 \pm
0.6\right)\times 10^3\;{\rm MeV}^2$.
}
\label{fig3}
\end{figure}
%
%end Figure 3
%

%
%begin Figure 4
%
\begin{figure}[hbt]
%\begin{center}
%\include{ps}
%\end{center}
\caption[]{The $\pi^+\pi^-$ phase shifts in the vector--isovector
channel calculated with the $\rho$--propagator (\ref{rppd}) which is dressed by
$\pi$--loops. Note that the $\omega$--meson does not contribute here.
The data (represented by circles) are taken from \cite{FP77}.
}
\label{fig4}
\end{figure}
%
%end Figure 4
%

Ignoring for the moment the $\rho\omega$--mixing but including
the pion loop the transversal part of the $\rho$ propagator
would have been given by
\beq
D_\rho^{ab}= \frac{\delta^{ab}}{p^2-m^2_\rho -\Sigma_\rho}\quad .
\label{rppd}
\eeq
Since we have ignored two pion loops and vector meson loops the
$\omega$--propagator does not receive a width. As the
$\rho\omega$--mixing is small the details of the $\omega$--propagator
will not be substantial for the $\rho$--propagator. Since we are
interested here in processes exclusively determined by the
$\rho$--propagator we will not attempt a detailed evaluation of the
$\omega$--propagator but rather use the empirical Breit--Wigner form
\beq
D_\omega= \frac{1}{p^2-m^2_\omega -i\Gamma_\omega m_\omega}\quad ,
\label{bwom}
\eeq
where $m_\omega$ and $\Gamma_\omega$ are the empirical mass and
width of the $\omega$--meson. Note that the Breit--Wigner form
ignores any momentum dependence of $\Gamma_\omega$ or $m_\omega$,
which would result in a two loop calculation from the
Wess--Zumino term. We will see however later that a possible
momentum dependence of the width $\Gamma_\omega$ or the mass
$m_\omega$ is completely irrelevant for the $\rho$--propagator
(since the $\rho \omega$--mixing is so small that it matters only
near the pole). Assuming the empirical Breit--Wigner shape
(\ref{bwom}) for
the $\omega$--propagator the $\rho\omega $--mixing modifies the
free $\rho$--propagator by the factor
\beq
\left(1+\frac{1}{3}\frac{m_\omega^2}{m_\rho^2}
\frac{m_{\rho\omega}^2}{p^2-m_\omega^2 + i m_\omega\Gamma_\omega}\right)\quad .
\eeq
This is immediately seen by diagonalizing the $\rho \omega$--mixing
after having the free $\omega$--lagrangian replaced by the
Breit--Wigner form
\beq
{\cal L}_\omega = \half\omega\left( \partial^2 + m_\omega^2
-im_\omega\Gamma_\omega \right)\omega\quad .
\eeq

Therefore we finally obtain for the $\rho$--propagator, which
includes both the pion loop as well as the $\rho \omega$--mixing,
the following expression
\beq
D_\rho^{ab}= \delta^{ab}\frac{1}{p^2-m^2_\rho -\Sigma_\rho}
\left(1+\frac{1}{3}\frac{m_\omega^2}{m_\rho^2}
\frac{m_{\rho\omega}^2}{p^2-m_\omega^2 + i m_\omega\Gamma_\omega}\right)\quad .
\eeq

Once the $\rho$ propagator is known the electromagnetic form
factor of the pion and the $\pi^+\pi^-$ phase shifts can be
evaluated from (\ref{ffem}) and (\ref{ps}).

\section{Numerical results}
In the actual numerical calculation
let us first consider the effective meson theory derived from the
bosonized NJL model in the low energy regime as a
phenomenological meson model. That is we determine the involved
parameters from the experimental meson data. Using the
experimental values $m_\omega = 782 {\rm MeV}$ and $\Gamma_\omega =
{\rm 843} MeV$ we can adjust $\Lambda_\pi, m_\rho$ and $g_{\rho\pi\pi}$ to
fix the mass and width of the $\rho$--meson and the height of the
electromagnetic form factor. Furthermore the $\rho \omega$--mixing
strength is adjusted to reproduce the fine structure of
the electromagnetic pion form factor.

This yields
\beq
\Lambda_\pi &=& 663\; {\rm MeV}\nonumber\\
m_\rho^0 &=& 922.5\;  {\rm MeV}\nonumber\\
g_{\rho\pi\pi} &=& 5.86\nonumber \\
m_{\rho\omega}^2 &=&-4850\; {\rm MeV}^2\nonumber
\eeq

The meson loop cutoff $\Lambda_\pi$ is in fact smaller than the
quark loop cutoff $\Lambda \sim 1.3\, {\rm MeV}$ as anti\-cipated. The
bare $\rho$--mass $m^0_\rho$ is somewhat larger than the physical
$\rho$--mass which is determined by the position of the peak in
the electromagnetic form factor. Amazingly the $\rho\pi\pi$
coupling constant obtained from the fit of the electromagnetic
form factor $g_{\rho\pi\pi} = 5.86$ agrees very well with the
value predicted from the bosonized NJL model $g_{\rho\pi\pi} =6 $.
The form factor obtained with these parameters is shown in figure
(\ref{fig2}).  It amazingly well reproduces
the experimental form factor. Note the kink in the form factor
on the right hand side above the peak. Our analysis shows that
this kink is exclusively generated by the $\rho \omega$--mixing.
This comes out also in the analysis of ref. \cite{SP92}.

Let us finally interpret the empirically determined value of
the $\rho \omega$--mixing in terms of the result from the
bosonized NJL model. Using in (\ref{mrw2}) the empirical value for the
mixing strength
and fixing all parameters of the NJL model except the
isospin breaking part $m_u - m_d$ in the standard fashion from
$F_\pi, m_\pi, g_{\rho\pi\pi}$ yields for the splitting between the up and
down current quark masses
\beq
m_d-m_u \approx (1-2)\,{\rm MeV}\quad ,
\eeq
which is not unrealistic.
The momentum dependence of the mixing strength $m_{\rho\omega}^2$ for
three different values of $\Delta \Sigma$ can be seen in figure
(\ref{fig3}). Notice that $m_{\rho\omega}^2$ vanishes for $p^2=0$
independent of the
mass splitting of the quarks as a consequence of gauge symmetry.

Finally with the above determined $\rho$ propagator we can now
calculate the $\pi^+\pi^-$ phase shift in the  vector--isovector channel
in a parameter free way. The result is shown in figure (\ref{fig4}).

The
theoretically curve excellently reproduces the experimental
data. Let us emphasize that our investigations show that the inclusion
of the pion loop is crucial for obtaining the correct energy
dependence of the phase shift. Note also that the $\rho
\omega$--mixing
does not contribute to the phase shift but is crucial to
obtain the fine structure of the electromagnetic form factor.


\begin{thebibliography}{99}
\bibitem{GL84} S.~Weinberg, Phys.\ Rev.\
               Lett.\ {\bf 18} (1967) 507;
J.~Gasser and H.~Leutwyler, Ann.~Phys. {\bf 158}
(1984) 142.
\bibitem{ER86}{D. Ebert and H. Reinhardt,
             Nucl. Phys. {\bf B271} (1986) 188.}

\bibitem{MNS90} G.~A.~Miller, B.~M.~K.~Nefkens and I.~Slaus,
                Phys.~Rep. {\bf 194} (1990) 1.

\bibitem{Vo86} M.~K.~Volkov,
                Sov.~J.~Part.~Nucl. {\bf 17} (1986) 186.

\bibitem{NJL61} Y. Nambu and G. Jona-Lasinio,
		Phys. Rev. {\bf D122} (1961) 345;
		ibid. {\bf D124} (1961) 246.


\bibitem{RD89} H. Reinhardt and B. V. Dang,
		Nucl. Phys. {\bf A500} (1989) 563.

\bibitem{Fu80} K.~Fujikawa, Phys.~Rev. {\bf D21} (1980) 2848; {\bf
D23} (1981) 2262; {\bf D29} (1984) 285.
\bibitem{WZ71} J. Wess and B. Zumino, Phys. Lett. {\bf 37B} (1971) 95.
\bibitem{BKY88}M. Bando, T. Kugo and K. Yamawaki,
                Phys. Rep. {\bf 164} (1988) 217.
\bibitem{SP92} J.~Speth and B.~C.~Pearce, {\it The structure of
Mesons and Nucleons}, proceedings of {\it 4th International Spring
Seminar on Nuclear Physics}, Amalfi, ed. A.~Covello, (1992) 29.
\bibitem{Am84} S.\ R.\ Amendolia et al., Phys.\ Lett.\ {\bf B138} (1984)
454;\\ S.\ R.\ Amendolia et al., Phys.\ Lett.\ {\bf B146} (1984)
116;\\ L.\ M.\ Barkov et al., Nucl.\ Phys.\ {\bf B256} (1985) 365.



\bibitem{FP77} C.~D.~Froggatt and J.~L.~Petersen,
                 Nucl.~Phys.~{\bf B129} (1977) 89.

\end{thebibliography}
\end{document}